# From wave-functions to current-voltage characteristics: overview of a Coulomb blockade device simulator using fundamental physical parameters.

Johann Sée, Philippe Dollfus, Sylvie Galdin, and Patrice Hesto

*Institut d'Électronique Fondamentale (CNRS UMR 8622),*

*Université Paris XI 91405 Orsay, France*

## Abstract

The purpose of this article is to present an accurate way, based on a physical description, to simulate Coulomb blockade devices. The method underlying the simulations depends only on fundamental parameters of the system and does not require the use of high level fitting parameters as tunneling conductances contrary to number of current Coulomb blockade simulators. It lies mainly on the transfer Hamiltonian formalism and Bardeen's formula within the framework of effective mass tensor. It can be applied to metallic Coulomb blockade devices as well as semiconductor ones. The details of this method are extensively reviewed from a theoretical point of view and the main results are presented. In particular, we study how to obtain tunneling rates information to deduce current/voltage characteristics of Metal—Insulator—Metal—Insulator—Metal (MIMIM) and Metal—Insulator—Si Quantum Dot—Insulator—Metal (MI*Si*IM) structures.

Keywords: Coulomb blockade, tunneling current, single electron devices

**Contents**

# I. INTRODUCTION

## A. New simulation challenges for new microelectronic device concepts

Among the new architecture concepts aiming at pursuing the increase in density imposed by the roadmap, the silicon single-electron devices appear to be potential candidates to improve, to complete or even to replace the current metal-oxide-semiconductor (MOS) technology with which they may remain compatible. Indeed, the use of the Coulomb blockade phenomenon in systems made up of combinations of tunnel junctions and semiconductor quantum dots, seems to offer promise perspectives in particular in non-volatile memory applications and also for single-electron transistor applications[1]. Thus, the concept of multi-dot memory using silicon nanocrystals embedded in silicon dioxide as floating-gate has already been demonstrated[2] and the quantization effects have been used in self-aligned double-stacked memory to improve the retention time[3].

Within this context, the simulations of such devices must be performed not only to understand but also to predict experimental behaviors. Moreover, from a physical point of view we will learn a lot from these simulations if they are independent on high-level experimental parameters (as tunneling rates) and based on low-level "concrete" ones (geometrical data, barrier height,...). In previous articles[4,5] we have already presented a set of models devoted to the electronic structure calculation in a semiconductor quantum dot embedded in insulator. Here we use these results to model the electron transport through tunnel barriers. Thus, after a review of the tunneling transfer Hamiltonian method, we develop an extension of Bardeen's formula[6], in the framework of effective mass tensor, which can be used to determine tunneling rates from the electron wave-functions. In particular, the tunneling rates of metallic and semiconductor Coulomb blockade devices containing one island are examined. Finally $I(V)$ characteristics are calculated with a Monte-Carlo technique.

## B. Transfer Hamiltonian: a first approach

A usual method employed to describe the tunneling processes in Coulomb blockade devices is based on tunneling Hamiltonian approach. This theory, thoroughly studied by many authors[7,8,9], aims at treating the tunneling events as a perturbation (provided the transmission coefficient $T$ of the barrier is small: $T \ll 1$). Thus, the Hamiltonian $\mathcal{H}$ of a single barrier is decomposed in three

Hamiltonians as follows (cf. Fig. 1):

$$\mathcal{H} = \mathcal{H}_L + \mathcal{H}_R + \mathcal{H}_T, \tag{1}$$

where $\mathcal{H}_L$ and $\mathcal{H}_R$ represent the Hamiltonians of the left and right electrodes, respectively, as if they were "alone" and independent. The Hamiltonian $\mathcal{H}_T$ expresses the perturbation induced by the tunnel process. One of the most attractive advantage in the use of a perturbation technique is linked to the capability of treating complex system, many-body effects and interactions between electrons, which is fundamental in Coulomb blockade devices.

The second quantization is said to offer a natural and transparent way to study the significance of the tunneling Hamiltonian. Indeed this one can be written in this formalism:

$$\mathcal{H} = \mathcal{H}_L + \mathcal{H}_R + \mathcal{H}_T = \sum_{\vec{k}_L} \mathcal{E}^L_{\vec{k}_L} a^\dagger_{\vec{k}_L} a_{\vec{k}_L} + \sum_{\vec{k}_R} \mathcal{E}^R_{\vec{k}_R} b^\dagger_{\vec{k}_R} b_{\vec{k}_R} + \mathcal{H}_T, \tag{2}$$

with

$$\mathcal{H}_T = \sum_{\vec{k}_L, \vec{k}_R} T_{\vec{k}_L \vec{k}_R} b^\dagger_{\vec{k}_R} a_{\vec{k}_L} + \sum_{\vec{k}_R, \vec{k}_L} T_{\vec{k}_R \vec{k}_L} a^\dagger_{\vec{k}_L} b_{\vec{k}_R}. \tag{3}$$

Where $a^\dagger_{\vec{k}_L}$ and $a_{\vec{k}_L}$ (resp. $b^\dagger_{\vec{k}_R}$ and $b_{\vec{k}_R}$) are the fermion creation and annihilation operators of the left (resp. right) electrode sub-system (cf. Fig. 1) and $\mathcal{E}^L_{\vec{k}_L}$ (resp. $\mathcal{E}^R_{\vec{k}_R}$) represents the energy of the eigenstate $|k_R\rangle$ (resp. $|k_L\rangle$). The matrix coefficient $T_{\vec{k}_R \vec{k}_L} = \langle k_R | \mathcal{H}_T | k_L \rangle$ (resp. $T_{\vec{k}_L \vec{k}_R}$) quantifies the probability for a particle to transfer from a state of the left (resp. right) electrode to a state of the right (resp. left) electrode by a tunnel process. The tunneling rate is then determined *via* the time-dependent perturbation theory by taking the electron from one side of the barrier as initial state and the electron on the other side as final state. Using Fermi's golden rule[10,11] we can calculate the tunneling rate from a left state $|k_L\rangle$ (part of a continuum) to a right state $|k_R\rangle$ (part of a continuum):

$$\delta^2 \mathcal{P}_{\vec{k}_L \rightarrow \vec{k}_R} = \frac{2\pi}{\hbar} \left| \langle \vec{k}_R | \mathcal{H}_T | \vec{k}_L \rangle \right|^2 \delta(\mathcal{E}_R - \mathcal{E}_L) \rho_R(\mathcal{E}_R) \rho_L(\mathcal{E}_L) \, d\mathcal{E}_R d\mathcal{E}_L, \tag{4}$$

with $\rho_L$ (resp. $\rho_R$) the density of states of the left (resp. right) electrode. Introducing the fermion energy distribution for an equilibrium ground state *via* the Fermi-Dirac statistics

$$f_{R/L}(\mathcal{E}) = f(\mathcal{E} - E_{FR/L}) = \frac{1}{1 + \exp\left(\frac{\mathcal{E} - E_{FR/L}}{k_B T}\right)}, \tag{5}$$

we can evaluate the total tunneling rate from all occupied states of the left to all unoccupied states on the right:

$$\Gamma_{L\to R} = \int\limits_{-\infty}^{+\infty} \rho_L(\mathcal{E}_L) f_L(\mathcal{E}) \left\{ \int\limits_{-\infty}^{+\infty} \frac{2\pi}{\hbar} \left|\left\langle \vec{k}_R \middle| \mathcal{H}_T \middle| \vec{k}_L \right\rangle\right|^2 \delta(\mathcal{E}_R - \mathcal{E}_L) \left[1 - f_R(\mathcal{E})\right] \rho_R(\mathcal{E}_R) \, d\mathcal{E}_R \right\} d\mathcal{E}_L. \tag{6}$$

In this equation a fundamental hidden parameter is given by the Fermi energy $E_{FR/L}$. Indeed, linked to the chemical potential notion, the relation between the Fermi level of the electrodes can be written $E_{F_R} = E_{F_L} + \Delta\mathcal{F}/\Delta N$, where the left electrode is considered as the reference and $\Delta\mathcal{F}$ is the increase in free energy due to the tunneling of $\Delta N$ particles. For a simple barrier where electron electrostatic interaction is negligible, $\Delta\mathcal{F}/\Delta N$ is equal to the bias potential applied $-qV$.

From now, several more or less sophisticated models[12,13] could be employed to simplify Eq. (6). In particular, considering as constants the transmission matrix element and the densities of states along the range of energy in the integration leads to the concept of tunneling conductance. Indeed the tunneling current $I = -q[\Gamma_{L\to R} - \Gamma_{R\to L}]$ becomes proportional to the bias potential applied on the barrier: $I = VG_t$ with

$$G_t = \frac{2\pi q^2 \rho_L \rho_R |T|^2}{\hbar}. \tag{7}$$

The counterpart of this simple and naive description of the tunneling transfer Hamiltonian lies in the fact that several hypothesis are not mentioned and that second quantized form (2) is only available on specific conditions which contradict Coulomb blockade phenomenon. First, the decomposition of equation (2) is wrong if we take into account electrostatic interaction between the electrons since we must introduce a two-body operator. Especially, when studying semiconductor quantum dot, the Hamiltonian of an island exhibiting Coulomb blockade phenomenon must be written

$$\mathcal{H}_{\text{island}} = \sum_k \mathcal{E}_k a_k^\dagger a_k + \sum_{klmn} \mathcal{O}_{klmn} a_k^\dagger a_l^\dagger a_m a_n, \tag{8}$$

where $\mathcal{H}_{\text{island}}$ is decomposed on the basis of the eigenstates of the island with no electrostatic interaction and $\mathcal{O}_{klmn}$ represents the electron-electron Coulombian interaction.

Nevertheless, even for metallic island where the electrostatic interactions are taken into account *via* Fermi level evolution, the second quantized form (2) underlies some assumptions. In particular,

- the eigenstates $\{|\Phi_L^n\rangle\}_n$ of $\mathcal{H}_L$ must form a complete basis of $\mathcal{H}_L$ ;

- likewise, the eigenstates $\{|\Phi_R^m\rangle\}_m$ of $\mathcal{H}_R$ must form a complete basis of $\mathcal{H}_R$ ;

- to decompose the non-perturbed Hamiltonian as $\sum_{\vec{k}_L} \mathcal{E}^L_{\vec{k}_L} a^\dagger_{\vec{k}_L} a_{\vec{k}_L} + \sum_{\vec{k}_R} \mathcal{E}^R_{\vec{k}_R} b^\dagger_{\vec{k}_R} b_{\vec{k}_R}$, the eigenstates $\{|\Phi^n_L\rangle\}_n$ and $\{|\Phi^m_R\rangle\}_m$ must be independent, that is to say all eigenstates of $\mathcal{H}_L$ must be orthogonal to any eigenstates of $\mathcal{H}_R$ ;

- and $\{|\Phi^n_L\rangle\}_n \bigcup \{|\Phi^m_R\rangle\}_m$ must form a complete basis for the total Hamiltonian $\mathcal{H}$.

This point has already been studied by Prange and other authors[14,15] and leads to the conclusion that no physical states can fulfil the four above conditions. One solution consists in imposing the completeness to the detriment of the orthogonality: the states are therefore taken as orthogonal as possible.

Finally, second quantized form does not offer any way to evaluate the transmission matrix element $T_{\vec{k}_L \vec{k}_R} = \langle \vec{k}_R | \mathcal{H}_T | \vec{k}_L \rangle$. That is why many Coulomb blockade simulators use the tunneling conductance as a fitting parameter. Nevertheless, this "high-level" concept must be manipulated with a great care, especially when studying semiconductor devices where quantization and interaction phenomena are far more complex than in metallic cases.

To circumvent this last issue, Bardeen's formula[6,16] could be used to determine $T_{\vec{k}_L \vec{k}_R}$. Nevertheless, it is necessary to show to what extent this formula could be used to describe semiconductor quantum dots embedded in silicon dioxide and to enlighten the hypothesis underlying it. From this point of view, we detail in the next section the calculation of the tunneling rates from the decomposition of the barrier Hamiltonian $\mathcal{H}$ within the framework of effective mass tensor. The demonstration will be available for any kind of Coulomb blockade devices either metallic or semiconductor.

## II. FROM WAVE-FUNCTIONS TO TUNNELING RATES BY THE TRANSFER HAMILTONIAN METHOD

### A. Transfer Hamiltonian theory in the mass tensor formalism

We write the Hamiltonian of the barrier as

$$\mathcal{H} = -\frac{\hbar^2}{2}\vec{\nabla}[M]^{-1}\vec{\nabla}. + V(\vec{r}) \quad \text{with} \quad \mathcal{H}\psi = E\psi, \quad (9)$$

and, for the decomposition, we define the left and right electrode[22] Hamiltonian by

$$\mathcal{H}_L = -\frac{\hbar^2}{2}\vec{\nabla}[M_L]^{-1}\vec{\nabla}. + V_L(\vec{r}) \quad \text{with} \quad \mathcal{H}_L\Phi_L^m = E_L^m\Phi_L^m, \tag{10}$$

$$\mathcal{H}_R = -\frac{\hbar^2}{2}\vec{\nabla}[M_R]^{-1}\vec{\nabla}. + V_R(\vec{r}) \quad \text{with} \quad \mathcal{H}_R\Phi_R^n = E_R^n\Phi_R^n, \tag{11}$$

where $[M]$, $[M_L]$ and $[M_R]$ represent the electron effective mass tensors. In order to decompose the barrier Hamiltonian we will introduce two domain functions[8] $\Theta_L$ and $\Theta_R$ delimiting each electrode. By definition, the domain function $\Theta_L$ (resp. $\Theta_R$) is equal to unity on the left (resp. right) domain and is equal to zero elsewhere. Moreover, these functions are built so that $\Theta_L \cdot \Theta_R = 0$ (there is no intersection between the domains) and $\Theta_L + \Theta_R = 1$ (the union of the domains represents the whole space). The choice of the delimiting area between the electrodes is arbitrary but a good one consists in taking the frontier in the middle of the barrier.

The potential $V(\vec{r})$ can be easily decomposed if we choose the value of $V_L(\vec{r})$ (resp. $V_R(\vec{r})$) equal to $V(\vec{r})$ on the left (resp. right) domain and take a constant in the right (resp. left) domain:

$$\begin{cases} V_L(\vec{r}) = \Theta_L V(\vec{r}) + \Theta_R V_0 \\ V_R(\vec{r}) = \Theta_R V(\vec{r}) + \Theta_L V_0 \end{cases}, \tag{12}$$

where the constant $V_0$ (identical for $V_L(\vec{r})$ and $V_R(\vec{r})$ ) corresponds to a value of the potential at the frontier between the two electrode domains. Likewise, we can define for effective mass tensor:

$$\begin{cases} [M_L](\vec{r}) = \Theta_L[M](\vec{r}) + \Theta_L[M_0] \\ [M_R](\vec{r}) = \Theta_R[M](\vec{r}) + \Theta_R[M_0] \end{cases}, \tag{13}$$

where $[M_0]$ is the mass tensor in the tunnel barrier. With these notations, we can write the following relations between $\mathcal{H}$, $\mathcal{H}_L$ and $\mathcal{H}_R$

$$\mathcal{H} = \Theta_L\mathcal{H}_L + \Theta_R\mathcal{H}_R, \tag{14}$$

$$\mathcal{H}_L = \Theta_L\mathcal{H}_L + \Theta_R\left[-\frac{\hbar^2}{2}\vec{\nabla}[M_0]^{-1}\vec{\nabla}. + V_0\right] \tag{15}$$

$$\mathcal{H}_R = \Theta_R\mathcal{H}_R + \Theta_L\left[-\frac{\hbar^2}{2}\vec{\nabla}[M_0]^{-1}\vec{\nabla}. + V_0\right]. \tag{16}$$

Now, to calculate the tunneling rates we have to solve the time-dependent Schrödinger equation

$$i\hbar\frac{d}{dt}|\psi\rangle = \mathcal{H}|\psi\rangle, \tag{17}$$

knowing that at $t = 0$, $|\psi\rangle$ is an eigenstate $|\Phi_L^0\rangle$ of the left electrode Hamiltonian and that after the tunnel event, we search for the probability that $|\psi\rangle$ is an eigenstate of the right electrode Hamiltonian. We treat the problem in the same way as we can prove Fermi's golden rule[10,11]: that is we search $|\psi\rangle$ in the form

$$|\psi\rangle = a(t)|\Phi_L^0\rangle e^{-iE_L^0 t/\hbar} + \sum_n b_n(t)|\Phi_R^n\rangle e^{-iE_R^n t/\hbar}, \qquad (18)$$

coupled to the initial conditions

$$a(t=0) = 1 \text{ and } \forall n \in \mathbb{N} \quad b_n(t=0) = 0. \qquad (19)$$

We can then replace the above expression of $|\psi\rangle$ in the time-dependent Schrödinger equation where we have decomposed the barrier hamiltonian $\mathcal{H}$ into $\mathcal{H} = \Theta_L \mathcal{H}_L + \Theta_R \mathcal{H}_R$.

$$\left[i\hbar \frac{da(t)}{dt} + E_L^0 a(t)\right] e^{-iE_L^0 t/\hbar} |\Phi_L^0\rangle + \sum_n \left[i\hbar \frac{db_n(t)}{dt} + E_R^n b_n(t)\right] e^{-iE_R^n t/\hbar} |\Phi_R^n\rangle =$$
$$a(t)\Theta_L \mathcal{H}_L |\Phi_L^0\rangle e^{-iE_L^0 t/\hbar} + \sum_n b_n(t)\Theta_R \mathcal{H}_R |\Phi_R^n\rangle e^{-iE_L^0 t/\hbar} \qquad (20)$$
$$+ \sum_n b_n(t)\Theta_L \mathcal{H}_L |\Phi_R^n\rangle e^{-iE_R^n t/\hbar} + a(t)\Theta_R \mathcal{H}_R |\Phi_L^0\rangle e^{-iE_L^0 t/\hbar}$$

To go further, we substitute $\Theta_R \mathcal{H}_R$ and $\Theta_L \mathcal{H}_L$ thanks to the relations (16) and (15) and we use the definition of the eigenstates of the left and right Hamiltonians $\mathcal{H}_L |\Phi_L^0\rangle = E_L^0 |\Phi_L^0\rangle$ and $\mathcal{H}_R |\Phi_R^n\rangle = E_R^n |\Phi_R^n\rangle$. The equation (20) can then be simplified:

$$i\hbar \frac{da(t)}{dt} e^{-iE_L^0 t/\hbar} |\Phi_L^0\rangle + \sum_n i\hbar \frac{db_n(t)}{dt} e^{-iE_R^n t/\hbar} |\Phi_R^n\rangle =$$
$$a(t)\Theta_R \left\{\mathcal{H}_R - \left[-\frac{\hbar^2}{2}\vec{\nabla}[M_0]^{-1}\vec{\nabla}. + V_0\right]\right\} |\Phi_L^0\rangle e^{-iE_L^0 t/\hbar} \qquad (21)$$
$$+ \sum_n b_n(t)\Theta_L \left\{\mathcal{H}_L - \left[-\frac{\hbar^2}{2}\vec{\nabla}[M_0]^{-1}\vec{\nabla}. + V_0\right]\right\} |\Phi_R^n\rangle e^{-iE_R^n t/\hbar}$$

First, before studying the above equation with the perturbation theory, we set as hypothesis that the eigenstates $|\Phi_L^0\rangle$ and $|\Phi_R^n\rangle$ are quasi-orthogonal, i.e. $\langle \Phi_L^0 | \Phi_R^n \rangle \approx 0$. Strictly speaking, we have already said in the introduction that we cannot suppose that $|\Phi_L^0\rangle$ and $|\Phi_R^n\rangle$ are orthogonal and form at the same time a complete basis for $\mathcal{H}$. Nevertheless, we could suppose that $\langle \Phi_L^0 | \Phi_R^n \rangle$ is a second order term[23] (first order term being the tunnel event perturbation) and is negligible if we study the Schrödinger equation (21) at first order. Thus, after a projection on $\langle \Phi_L^0 |$, neglecting

term of order greater than one and using the fact that $\{|\Phi_L\rangle\}$ is an orthonormal basis, we find

$$i\hbar \frac{da(t)}{dt} e^{-iE_L^0 t/\hbar} = a(t) \left\langle \Phi_L^0 \left| \Theta_R \left\{ \mathcal{H}_R - \left[ -\frac{\hbar^2}{2}\vec{\nabla}[M_0]^{-1}\vec{\nabla}. + V_0 \right] \right\} \right| \Phi_L^0 \right\rangle e^{-iE_L^0 t/\hbar}$$

$$+ \underbrace{\sum_n b_n(t) \left\langle \Phi_L^0 \left| \Theta_L \left\{ \mathcal{H}_L - \left[ -\frac{\hbar^2}{2}\vec{\nabla}[M_0]^{-1}\vec{\nabla}. + V_0 \right] \right\} \right| \Phi_R^n \right\rangle e^{-iE_R^n t/\hbar}}_{\text{tunnel transfer term}}.$$

(22)

A projection on $\langle \Phi_R^k |$ gives:

$$i\hbar \frac{db_k(t)}{dt} e^{-iE_R^k t/\hbar} = \underbrace{a(t) \left\langle \Phi_R^k \left| \Theta_R \left\{ \mathcal{H}_R - \left[ -\frac{\hbar^2}{2}\vec{\nabla}[M_0]^{-1}\vec{\nabla}. + V_0 \right] \right\} \right| \Phi_L^0 \right\rangle e^{-iE_L^0 t/\hbar}}_{\text{tunnel transfer term}}$$

$$+ \sum_n b_n(t) \left\langle \Phi_R^k \left| \Theta_L \left\{ \mathcal{H}_L - \left[ -\frac{\hbar^2}{2}\vec{\nabla}[M_0]^{-1}\vec{\nabla}. + V_0 \right] \right\} \right| \Phi_R^n \right\rangle e^{-iE_R^n t/\hbar}.$$

(23)

It should be noted that the preceding equations are constituted by two matrix element terms. One of this matrix element is directly related to the tunnel effect since it quantifies the probability for an initial left electrode state to transfer in a right electrode state (or *vice-versa*). This matrix element must be considered as a first order term in the perturbation theory. The second matrix element term corresponds to the influence of one electrode on the eigenstates of the other electrode.

In perturbation theory, the coefficients $a(t)$ and $b_n(t)$ are decomposed in power of $\lambda$ ($a(t) = a^0(t) + \lambda a^1(t) + o(\lambda)$ and $b_n(t) = b_n^0(t) + \lambda b_n^1(t) + o(\lambda)$) and the matrix elements of the kind $\langle \psi | H | \varphi \rangle$ are replaced by $\lambda \langle \psi | H | \varphi \rangle$. Using the initial conditions we are led to conclude that $a^0 = \text{const} = 1$, $b_k^0 = \text{const} = 0$. The first order study in $\lambda$ gives:

$$i\hbar \frac{db_k^1}{dt} e^{-iE_R^k t/\hbar} = \left\langle \Phi_R^k \left| \Theta_R \left\{ \mathcal{H}_R - \left[ -\frac{\hbar^2}{2}\vec{\nabla}[M_0]^{-1}\vec{\nabla}. + V_0 \right] \right\} \right| \Phi_L^0 \right\rangle e^{-iE_L^0 t/\hbar}. \quad (24)$$

In fact, we find the same equations as in Fermi's golden rule with a time-independent perturbation Hamiltonian

$$\mathcal{H}_{\text{pert}} = \Theta_R \left\{ \mathcal{H}_R - \left[ -\frac{\hbar^2}{2}\vec{\nabla}[M_0]^{-1}\vec{\nabla}. + V_0 \right] \right\}. \quad (25)$$

Knowing that $\langle \Phi_R^k | \mathcal{H}_{\text{pert}} | \Phi_L^0 \rangle$ is time-independent and within the framework of short-time approximation (cf. Ref. 11), the integration of differential equation (24) gives

$$\left| b_k^1 \right|^2 = \frac{2\pi}{\hbar} t \left| \langle \Phi_R^k | \mathcal{H}_{\text{pert}} | \Phi_L^0 \rangle \right|^2 \delta \left( E_G^0 - E_R^k \right). \quad (26)$$

We are now able to determine the probability $P_{i \to f}$ to find a particle which has left the left electrode initial state $|\Phi_L^0\rangle$ to reach a final right electrode state $|\Phi_R^k\rangle$ at time $t$. This probability is defined by $P_{i \to f} = \langle \Phi_L^0 | \psi \rangle$:

$$P_{i \to f} = \left| a(t) e^{-i E_G^0 t/\hbar} \langle \Phi_D^k | \Phi_G^0 \rangle + \sum_n b_n(t) e^{-i E_D^n t/\hbar} \langle \Phi_D^k | \Phi_D^n \rangle \right|^2. \tag{27}$$

That is with our assumptions:

$$P_{i \to f} \approx \left| b_k^1 \right|^2. \tag{28}$$

As mentioned in the beginning of the article, the tunneling process can be modeled by Fermi's golden rule for a transition from a discrete initial left electrode state to a discrete right electrode final state due to a perturbation Hamiltonian $\mathcal{H}_{\text{pert}}$

$$P_{i \to f} = \frac{2\pi}{\hbar} t \left| \langle \Phi_D^k | \mathcal{H}_{\text{pert}} | \Phi_G^0 \rangle \right|^2 \delta\left( E_G^0 - E_D^k \right). \tag{29}$$

Nevertheless the obtention of this tunneling transfer probability is restricted by to some constraints:

- to apply the perturbation theory we must verify that the tunneling transfer probability is small ($P_{i \to f} \ll 1$) and that the equation (29) is limited to short-time study;

- less evident to satisfy *a priori*, the basis formed by $\{|\Phi_L\rangle\}$ and $\{|\Phi_R\rangle\}$ must be quasi-orthogonal, i.e. $\langle \Phi_L | \Phi_R \rangle = o(\lambda)$.

Finally, the tunneling rates $\gamma = dP_{i \to f}/dt$ corresponding to the probability of transfer per unit time is given by

$$\gamma = \frac{2\pi}{\hbar} \left| \langle \Phi_D^k | \mathcal{H}_{\text{pert}} | \Phi_G^0 \rangle \right|^2 \delta\left( E_G^0 - E_D^k \right). \tag{30}$$

The major improvement lies in the fact that we are now able to determine the matrix element $T_{\vec{k}_R \vec{k}_L} = M = \langle \Phi_R^k | \mathcal{H}_{\text{pert}} | \Phi_L^0 \rangle$. Indeed, we will show in the next section that we can calculate this coefficient, providing we know the electron wave-function in the two electrodes.

### B. Matrix element calculation: Bardeen's formula extension

By definition,

$$M = \langle \Phi_R^k | \mathcal{H}_{\text{pert}} | \Phi_L^0 \rangle = \iiint \overline{\Phi}_R^k(\vec{r}) \mathcal{H}_{\text{pert}} \Phi_L^0(\vec{r}) \, d^3\vec{r}. \tag{31}$$

The consequence of the $\Theta_R$ function in the $\mathcal{H}_{\text{pert}}$ definition of expression (25) is the limitation of the integration domain over the right one only

$$M = -\iiint\limits_{\text{right}} \overline{\Phi}_R^k(\vec{r}) \left[ -\frac{\hbar^2}{2} \vec{\nabla}[M_0]^{-1}\vec{\nabla}. + V_0(\vec{r}) \right] \Phi_L^0(\vec{r})\, d^3\vec{r}$$
$$+ \iiint\limits_{\text{right}} \overline{\Phi}_R^k(\vec{r}) \left[ -\frac{\hbar^2}{2} \vec{\nabla}[M_R]^{-1}\vec{\nabla}. + V_R(\vec{r}) \right] \Phi_L^0(\vec{r})\, d^3\vec{r}. \quad (32)$$

Now, the right domain corresponds to the "barrier part" of the left electrode. As a consequence, *in the right domain* (and only in this one), $-\hbar^2/2\vec{\nabla}[M_0]^{-1}\vec{\nabla}. + V_0$ corresponds to $\mathcal{H}_L$. We can then write that in *the right domain* $\left[ -\hbar^2/2\vec{\nabla}[M_0]^{-1}\vec{\nabla}. + V_0 \right] |\Phi_L^0\rangle = E_L^0 |\Phi_L^0\rangle$. Thus, the equation (32) becomes:

$$M = -E_L^0 \left\langle \Phi_R^k \middle| \Phi_L^0 \right\rangle + \iiint\limits_{\text{right}} \overline{\Phi}_R^k \left[ -\frac{\hbar^2}{2} \vec{\nabla}[M_R]^{-1}\vec{\nabla}\Phi_L^0 \right] d^3\vec{r} + \iiint\limits_{\text{right}} \overline{\Phi}_R^k V_R(\vec{r})\Phi_L^0\, d^3\vec{r}. \quad (33)$$

To go further, we apply Green's formula[24] to the two following integrals:

$$\iiint\limits_{\text{right}} \overline{\Phi}_R^k \left[ -\frac{\hbar^2}{2} \vec{\nabla}[M_R]^{-1}\vec{\nabla}\Phi_L^0 \right] d^3\vec{r} = -\frac{\hbar^2}{2} \oiint\limits_{\mathcal{S}} \overline{\Phi}_R^k [M_R]^{-1}\vec{\nabla}\Phi_L^0\, d\vec{S}$$
$$+ \frac{\hbar^2}{2} \iiint\limits_{\text{right}} \vec{\nabla}\overline{\Phi}_R^k [M_R]^{-1}\vec{\nabla}\Phi_L^0\, d^3\vec{r} \quad (34)$$

$$\iiint\limits_{\text{right}} \Phi_L^0 \left[ -\frac{\hbar^2}{2} \vec{\nabla}[M_R]^{-1}\vec{\nabla}\overline{\Phi}_R^k \right] d^3\vec{r} = -\frac{\hbar^2}{2} \oiint\limits_{\mathcal{S}} \Phi_L^0 [M_D]^{-1}\vec{\nabla}\overline{\Phi}_R^k\, d\vec{S}$$
$$+ \frac{\hbar^2}{2} \iiint\limits_{\text{right}} \vec{\nabla}\Phi_L^0 [M_R]^{-1}\vec{\nabla}\overline{\Phi}_R^k\, d^3\vec{r} \quad (35)$$

In fact, the last terms of the two above equations are equal since they correspond to the two scalar products $\left( \vec{\nabla}\Phi_R^k \middle| [M_R]^{-1}\vec{\nabla}\Phi_L^0 \right)$ and $\left( [\overline{M_R}]^{-1}\vec{\nabla}\Phi_R^k \middle| \vec{\nabla}\Phi_L^0 \right)$. Indeed, thanks to the properties of the scalar product, we have

$$\left( \vec{\nabla}\Phi_R^k \middle| [M_R]^{-1}\vec{\nabla}\Phi_L^0 \right) = \left( [M_R]^{\dagger^{-1}}\vec{\nabla}\Phi_R^k \middle| \vec{\nabla}\Phi_L^0 \right) = \left( [\overline{M_R}]^{-1}\vec{\nabla}\Phi_R^k \middle| \vec{\nabla}\Phi_L^0 \right), \quad (36)$$

since we know that the effective mass tensor is a real and symmetrical tensor. As a result, we can rewrite the two equations (34) and (35):

$$\iiint_{\text{right}} \overline{\Phi}_R^k \left[ -\frac{\hbar^2}{2} \vec{\nabla} [M_R]^{-1} \vec{\nabla} \Phi_L^0 \right] d^3\vec{r} = -\frac{\hbar^2}{2} \oiint_S \overline{\Phi}_R^k [M_R]^{-1} \vec{\nabla} \Phi_L^0 \, d\vec{S}$$
$$+ \iiint_{\text{right}} \Phi_L^0 \left[ -\frac{\hbar^2}{2} \vec{\nabla} [M_R]^{-1} \vec{\nabla} \overline{\Phi}_R^k \right] d^3\vec{r} \quad (37)$$
$$+ \frac{\hbar^2}{2} \oiint_S \Phi_L^0 [M_R]^{-1} \vec{\nabla} \overline{\Phi}_R^k \, d\vec{S}.$$

Substitute the integral (37) in the equation (33) gives as new expression for the matrix element $M$:

$$M = -E_L^0 \left\langle \Phi_R^k \middle| \Phi_L^0 \right\rangle - \frac{\hbar^2}{2} \oiint_S \left[ \overline{\Phi}_R^k [M_R]^{-1} \vec{\nabla} \Phi_L^0 - \Phi_L^0 [M_R]^{-1} \vec{\nabla} \overline{\Phi}_R^k \right] d\vec{S}$$
$$+ \iiint_{\text{right}} \Phi_L^0 \left[ -\frac{\hbar^2}{2} \vec{\nabla} [M_R]^{-1} \vec{\nabla}. + V_R(\vec{r}) \right] \overline{\Phi}_R^k \, d^3\vec{r}. \quad (38)$$

In the right domain $\mathcal{H}_R = -(\hbar^2/2)\vec{\nabla}[M_R]^{-1}\vec{\nabla}. + V_R(\vec{r})$, and, we can say that on this domain $\left[ -\frac{\hbar^2}{2} \vec{\nabla}[M_R]^{-1}\vec{\nabla}. + V_R(\vec{r}) \right] \overline{\Phi}_R^k = E_R^k \overline{\Phi}_R^k$, so that the matrix element becomes

$$M = \left( E_R^k - E_L^0 \right) \left\langle \Phi_R^k \middle| \Phi_L^0 \right\rangle - \frac{\hbar^2}{2} \oiint_S \left[ \overline{\Phi}_R^k [M_R]^{-1} \vec{\nabla} \Phi_L^0 - \Phi_L^0 [M_R]^{-1} \vec{\nabla} \overline{\Phi}_R^k \right] d\vec{S}. \quad (39)$$

As we can see in the extension of Fermi's golden rule (29) the tunneling transfer probability is non zero only when $E_R^k = E_L^0$: the first term of the equation is then equal to zero[25]. In fact, as expected, we see that the tunneling effect occurs at constant energy, (even if we consider Coulomb blockade phenomenon).

The integration over a closed surface which leans on the whole right domain is useless since the area where particle can transfer is limited to the surface $\mathcal{S}_B$ of right domain which takes place inside the tunnel barrier. Moreover, on this surface, the effective mass tensor $[M_R]$ is constant and equal to $[M_0]$. Thus, the matrix element is finally given by:

$$M = \frac{\hbar^2}{2} \iint_{\mathcal{S}_B} \left[ \Phi_L^0 [M_0]^{-1} \vec{\nabla} \overline{\Phi}_R^k - \overline{\Phi}_R^k [M_0]^{-1} \vec{\nabla} \Phi_L^0 \right] d\vec{S}. \quad (40)$$

It should be noted that the relation between the tunneling matrix element corresponding to a transfer from a left state to a right state $M_{L \to R}$ and the tunneling matrix element corresponding to a transfer from a right state to a left state $M_{R \to L}$ is given by $M_{L \to R} = -\overline{M}_{R \to L}$. We find again that the tunneling probability (proportional to $|M|^2$) is independent of the direction of the transfer.

### C. The tunneling rates by transfer Hamiltonian method in a nutshell

Defining $\rho_{L/R}$ the density of states in the considered electrode, the various expressions of $\gamma$ depending on the electrode nature (continuum or discrete states) are given by:

- discrete states on the right and on the left:

$$\gamma = \frac{2\pi}{\hbar}|M|^2 \delta\left(E_L^0 - E_R^k\right) \tag{41}$$

- discrete case on the left and continuum on the right:

$$\gamma = \frac{2\pi}{\hbar}|M|^2 \delta\left(E_L^0 - E_R\right) \rho_R(E_R) dE_R \tag{42}$$

- continuum on the left and discrete states on the right:

$$\gamma = \frac{2\pi}{\hbar}|M|^2 \delta\left(E_L - E_R^k\right) \rho_L(E_L) dE_L \tag{43}$$

- continuum on the right and continuum on the left:

$$\gamma = \frac{2\pi}{\hbar}|M|^2 \delta\left(E_L - E_R\right) \rho_R(E_R) \rho_L(E_L) dE_L \, dE_R \tag{44}$$

In any case, the value of the matrix element $M$ remains unchanged and is given by equation (40).

Finally, the tunneling rates $\Gamma_{L \to R}$ in the case, for instance, of a continuum on the right and on the left is given by taking into account the electron energy distribution using the Fermi statistics:

$$\Gamma_{L \to R} = \int_{-\infty}^{+\infty} \int_{-\infty}^{+\infty} \frac{2\pi}{\hbar}|M|^2 \rho_R(E_R) \rho_L(E_L) \delta(E_L - E_R) \times f_L(E_L) \times \left[1 - f_R(E_R)\right] dE_L \, dE_R \tag{45}$$

## III. APPLICATION OF THE TUNNELING RATES DETERMINATION FOR THE COULOMB BLOCKADE DEVICES

### A. Tunneling conductance approximation

To evaluate the effectiveness of the tunneling transfer Hamiltonian method, a first test can be made by a comparison with an exact calculation[17] of the transmission coefficient. It should be also interesting to compare it with the tunneling conductance approximation since this method is widely used to describe Coulomb blockade devices. Indeed, in the particularly "simple" case of a one-dimensional tunnel barrier under bias voltage (that is two metallic electrodes separated by an insulator), we have an access to the exact analytical expression of the transmission coefficient (based on Airy functions) and, then, to the tunneling current.

The energetic potential $V(x)$ representing the biased one-dimensional Metal—Insulator–Metal (MIM) barrier is shown in figure 2: the potential inside the oxide is given by the linear relation

$$V(x) = V_1 + V_{\text{bar}} + \frac{qV}{d}\left(x + \frac{d}{2}\right), \tag{46}$$

where $V_1$ is the reference potential (in eV) in the $x < -d/2$ zone and $V_{\text{bar}}$ the barrier height (measured between the conduction band of the electrode and the insulator). For simplicity, the electron effective mass $m$ is supposed to be a constant along the device and equal to the electron mass $m_0$ in vacuum.

In the following section, we introduce the notation:

$$k_1 = \sqrt{\frac{2m}{\hbar^2}(E - V_1)} \quad \alpha = \sqrt{\frac{2m}{\hbar^2}(V_0 - E)} \quad \text{et} \quad k_3 = \sqrt{\frac{2m}{\hbar^2}(E - V_3)}. \tag{47}$$

with $V_0$, the value of the potential at $x = 0$ (cf Fig. 2) and $E$ the energy of the electron. The decomposition of the barrier Hamiltonian is represented in figure 2: the frontier between the right and left domains is chosen at $x = 0$ and, in a first step, the electrodes are supposed to be of finite length $L_L$ et $L_R$. These lengthes will tend to infinity at the end of the demonstration. Since we cannot find simple analytical expression of the eigenstates of the left and right electrodes, we use a WKB[18] approximation to express the wave-function $\psi(x)$ in the different part of the system

$$\psi(x) \approx \psi(0) e^{-\int_0^x \sqrt{\frac{2m}{\hbar^2}}\sqrt{V(x') - E}\, dx'}. \tag{48}$$

Solving the equation (48) on the left electrode gives:

$$\psi_L(x) = \begin{cases} A_L \sin\left[k_1(x + L_L + d/2)\right] \\ \text{if } -(L_L + d/2) \leqslant x < -d/2 \\ A_L \sin(k_1 L_L) \, e^{-\alpha \frac{2d}{3qV} \frac{1}{\sqrt{V_0 - E}} \left[\left(\frac{qV}{d}x + V_0 - E\right)^{3/2} - \left(-\frac{qV}{2} + V_0 - E\right)^{3/2}\right]} \\ \text{if } -d/2 \leqslant x < 0 \\ A_L \sin(k_1 L_L) \, e^{-\alpha \frac{2d}{3qV} \frac{1}{\sqrt{V_0 - E}} \left[(V_0 - E)^{3/2} - \left(-\frac{qV}{2} + V_0 - E\right)^{3/2}\right]} e^{-\alpha x} \\ \text{if } 0 \leqslant x \\ 0 \text{ elsewhere} \end{cases},$$

(49)

and the continuity of $\vec{\nabla} \psi_L$ at $x = -d/2$ involves

$$\sin^2(k_1 L_L) = \frac{k_1^2}{k_1^2 + \alpha^2 \left|1 - \frac{qV}{2(V_0 - E)}\right|}. \tag{50}$$

Moreover, normalization of the wave-function induces $|A_L|^2_{L_L \to \infty} = 2/L_L$ if the length of the electrode tends to infinity.

In the same way, we can find a similar expression for the right electrode if we substitute $x$ by $-x$, $V$ by $-V$ and $k_1$ by $k_3$.

Substituting the expressions of the wave-functions $\psi_L$ and $\psi_R$ in Bardeen's formula (40) the matrix element becomes in the simple case of a one-dimensional barrier:

$$|M|^2 = \left(\frac{\hbar^2}{2m}\right)^2 |A_L|^2 |A_R|^2 4\alpha^2 \sin^2(k_1 L_L) \sin^2(k_3 L_R) \\ \times e^{-2\alpha d \frac{2(V_0 - E)}{3qV} \left[\left(1 + \frac{qV}{2(V_0 - E)}\right)^{3/2} - \left(1 - \frac{qV}{2(V_0 - E)}\right)^{3/2}\right]} \tag{51}$$

We can now calculate the tunneling rates defined by equation (45) where the density of states in the electrode are equal to the classical relation $\rho_{L/R}(E) = L_{G/D}/(2\pi) \left(2m/\hbar^2\right)/k_{1/3}$:

$$\Gamma_{L \to R} = \int_{\max(E_{CL}, E_{CR})}^{+\infty} \frac{2\pi}{\hbar} |M|^2 \rho_R(\mathcal{E}) \rho_L(\mathcal{E}) f(\mathcal{E} - E_{FL}) \left[1 - f(\mathcal{E} - E_{FR})\right] d\mathcal{E}, \tag{52}$$

with $E_{CL/R}$ the position of the conduction band in the left/right electrode. In this last equation, we have used the property of the dirac distribution $\delta(E_D - E_G)$.

It should be noted that when one replaces $\sin^2(k_1 L_L)$, $\sin^2(k_3 L_R)$, $|A_{L/R}|^2$ and $\rho_{L/R}$ by their expressions, the tunneling rates become independent of $L_L$ and $L_R$. Moreover, the right electrode Fermi level $E_{FR}$ is related to the left electrode Fermi level $E_{FL}$ by the bias voltage applied $E_{FR} = E_{FL} + qV$.

Tunneling current is then deduced from the tunneling rates

$$I = -q\left[\Gamma_{L\to R} - \Gamma_{R\to L}\right]$$
$$= -q \int_{\max(E_{CL}, E_{CR})}^{+\infty} \frac{2\pi}{\hbar} |M|^2 \rho_R(\mathcal{E}) \rho_L(\mathcal{E}) \left[f_L(\mathcal{E}) - f_R(\mathcal{E})\right] d\mathcal{E}. \tag{53}$$

In the approximation of the tunneling conductance, the integration can be simplified and we find for the tunneling current:

$$I = V G_t \quad \text{with} \quad G_t = \frac{2\pi q^2 \rho_{R0} \rho_{L0} |M_0|^2}{\hbar}, \tag{54}$$

where $\rho_{L0/G0}$ and $M_0$ are respectively constants representing the density of states in the left/right electrode and the matrix element $M$ over the domain of integration. In this article we choose to take these constants equal to the value of the density of states (resp. the matrix element) at the energy $E_{FL}$.

The two graphs in figure 3 exhibit the current/voltage characteristics of a 12 Å and 20 Å Gold–SiO$_2$–Gold tunnel barrier where the height $V_{\text{bar}}$ is equal to 9.6 eV and the distance between the Fermi level and the conduction band is equal to 5.5 eV. We are led to conclude that the current obtained *via* the tunneling transfer Hamiltonian coupled to the WKB approximation (symbols) fits very well the current deduced from an exact calculation of the tunneling current in a triangular barrier (continuous line). This figure shows also the current obtained by the tunneling conductance approximation which appears to be a first order development of the current/voltage characteristic: the conductance represents the slope of the $I(V)$ curve at the origin.

Moreover, this approximation is all the worse that the barrier becomes thicker. This property can be explained by the fact that the magnitude of the variation of the transmission coefficient increases when the barrier thickness increases. Thus, for large barrier application, and especially in single-electron device simulation, the tunneling conductance is a very poor approximation.

### B. Metallic Coulomb blockade devices

A simple extension of the preceding calculations can be applied to simulate metallic single-electron devices. Indeed, in a device such as Metal—Insulator—Metallic Island—Insulator—Metal (cf. Fig. 4), where the energy quantization can be neglected, the four fundamental parameters, *i.e.* the tunneling rates, can be determined from equation (52). The Coulomb blockade phenomenon is taken into account as an additional effect which introduces an extra-electrostatic interaction energy in the Fermi energy of the island. Actually, when an electron tunnels from one electrode to the island its energy remains constant: only the electron which has the exact energy it will have in the island can transfer. So, the energy of the particle which transfers through the barrier does not evolve during the tunneling process. This is not the case of the Fermi energy in the island which changes under the effect of the change in the electrostatic interaction energy. The relation between the Fermi energy in the island $E_{F\,\text{dot}}$ and the one in the electrode $E_{F\,\text{elec}}$ (chosen as a reference) is given by:

$$E_{F\,\text{dot}} = E_{F\,\text{elec}} + \frac{\partial \mathcal{F}}{\partial N}, \tag{55}$$

where $\mathcal{F}$ is the free energy of the system and $N$ is the number of electrons stored in the island. In Coulomb blockade mechanism the electrons are said to tunnel sequentially. Assuming that they can only tunnel one by one, we have $\partial \mathcal{F}/\partial N \approx \mathcal{F}(N+1) - \mathcal{F}(N)$. For the case of a MIMIM structure the free energy $\mathcal{F}$ can be calculated by an energy balance method[7]:

$$\mathcal{F} = \frac{1}{2(C_L + C_R)} \left[ C_L C_R V^2 + (Nq)^2 \right] + \frac{qV}{C_L + C_R} (n_L C_R + n_R C_L), \tag{56}$$

with $C_L$ (resp. $C_R$) is the left (resp. right) electrode–island capacitance, $n_L$ is the number of electrons which have entered in the island from the left electrode and $n_R$ is the number of electrons which have left the island to go to the right electrode.

We are interested here in a parallelepiped electrode to derive an analytical expression of wavefunction and current. To study three-dimensional devices, the confining potential $V(x)$ in the electrodes (and the metallic island) is decomposed in three elementary potentials depending only on one space coordinate $V_x(x)$, $V_y(y)$ and $V_z(z)$: $V_x(x)$ corresponds to the potential described in the preceding subsection which takes into accounts the bias voltage applied between the island and the electrode ; $V_y(y)$ (resp. $V_z(z)$) is chosen so that $V_y(y) = 0$ (resp. $V_z(z) = 0$) in the metal (i.e. if $-L_y/2 < y < L_y/2$) and $V_y(y) = \infty$ (resp. $V_z(z) = \infty$) in the oxide[26]. The quantization induced by this infinite quantum well representation is suppressed by tending $L_y$ (resp. $L_z$) to infinity. To improve the description of the system, the electron effective mass is supposed to be equal to $m_0$ in the metal and equal to $m_{ox}$ in the oxide (for instance $m_{ox} = 0.5 m_0$ in silicon dioxide[19]).

Due to the effective mass variation, the total Hamiltonian cannot be decomposed in three elementary Hamiltonians $\mathcal{H}_x$, $\mathcal{H}_y$ and $\mathcal{H}_z$ depending only of one space coordinate, even so we suppose that the wave-function can be written[27] $\Psi = \psi_x(x)\psi_y(y)\psi_z(z)$. The $\psi_x(x)$ part has already been calculated in equations (49) except that the electron effective mass is no more a constant. The wave-function $\psi_y(y)$ (resp. $\psi_z(z)$) is approached by a plane-wave, since $L_y$ (resp. $L_z$) tends to infinity. With no more information about the electron properties along $y$ (resp. $z$) axis, we are led to impose the value of the wave vector $k_y$ (resp. $k_z$) of the plane-wave: our choice is to take $k_y = k_z = 0$, that is to neglect the electron momentum in these directions.

Under these assumptions, we find:

$$\Psi_L = \begin{cases} 0 & \text{if } x \leqslant -(L_{xL} + d/2) \text{ or } |y| > L_{y_L}/2 \text{ or } |z| > L_{z_L}/2 \\ A_L \sin\left[k_1(x + L_{xL} + d/2)\right] & \text{if } -(L_{xL} + d/2) \leqslant x < -d/2 \text{ and } |y| \leqslant L_{y_L}/2 \text{ and } |z| \leqslant L_{z_L}/2 \\ A_L \sin(k_1 L_{xL}) e^{-\alpha \frac{2}{3qE} \frac{1}{\sqrt{V_0 - \mathcal{E}}} \left[(qEx + V_0 - \mathcal{E})^{3/2} - \left(-\frac{qEd}{2} + V_0 - \mathcal{E}\right)^{3/2}\right]} & \text{if } -d/2 \leqslant x < 0 \text{ and } |y| \leqslant L_{y_L}/2 \text{ and } |z| \leqslant L_{z_L}/2 \\ A_L \sin(k_1 L_{xL}) e^{-\alpha \frac{2}{3qE} \frac{1}{\sqrt{V_0 - \mathcal{E}}} \left[(V_0 - \mathcal{E})^{3/2} - \left(-\frac{qEd}{2} + V_0 - \mathcal{E}\right)^{3/2}\right]} e^{-\alpha x} & \text{if } x \geqslant 0 \text{ and } |y| \leqslant L_{y_L}/2 \text{ and } |z| \leqslant L_{z_L}/2 \end{cases}$$
(57)

and a similar expression for the right wave-function where $x$ becomes $-x$, $E$ becomes $-E$ and $k_1$ becomes $k_3$. Moreover, we have

$$k_1 = \sqrt{\frac{2m_0}{\hbar^2}(\mathcal{E} - V_1)} \quad \text{et} \quad \alpha = \sqrt{\frac{2m_{ox}}{\hbar^2}(V_0 - \mathcal{E})} \tag{58}$$

$$\sin^2(k_1 L_{xL}) = \frac{k_1^2}{k_1^2 + \left(\frac{m_0}{m_{ox}}\right)^2 \alpha^2 \left|1 - \frac{qEd}{2(V_0 - \mathcal{E})}\right|} \tag{59}$$

$$|A_L|^2 = \frac{2}{L_{xL} L_{y_L} L_{z_L}} = \frac{2}{\mathcal{V}_L}, \tag{60}$$

and

$$k_3 = \sqrt{\frac{2m_0}{\hbar^2}(\mathcal{E} - V_3)} \quad \text{et} \quad \alpha = \sqrt{\frac{2m_{\text{ox}}}{\hbar^2}(V_0 - \mathcal{E})} \tag{61}$$

$$\sin^2(k_3 L_{x_R}) = \frac{k_3^2}{k_3^2 + \left(\frac{m_0}{m_{\text{ox}}}\right)^2 \alpha^2 \left|1 + \frac{qEd}{2(V_0 - \mathcal{E})}\right|} \tag{62}$$

$$|A_R|^2 = \frac{2}{L_{x_R} L_{y_R} L_{z_R}} = \frac{2}{\mathcal{V}_R}, \tag{63}$$

where $E$ represents the electric field inside the oxide barrier. We are then able to calculate the matrix element $M$ *via* Bardeen's Formula and the tunneling rates. It should be noted that for a three-dimensional system the density of states is now expressed as:

$$\rho(\mathcal{E}) = 2 \times \frac{L_x L_y L_z}{(2\pi)^2} \left(\frac{2m_0}{\hbar^2}\right)^{3/2} \sqrt{\mathcal{E} - E_c}. \tag{64}$$

Typical results of tunneling rates as a function of the bias voltage applied along a Al—SiO$_2$—Al—SiO$_2$—Al structure are presented in figure 5: the left barrier is 23 Å thick, the right one is 17 Å thick, the island has a surface $\mathcal{S} = L_y L_z$ equal to 15 nm$^2$, the Al work function is taken equal to 4.1 eV, the SiO$_2$ electronic affinity to 0.9 eV and the Aluminium Fermi energy height to 11.6 eV. To determine the various capacitances of the system, a simple planar-capacitor expression is used but the capacitances can be calculated more accurately thanks to a finite-element method for instance.

In figure 6, we present a comparison between a full calculation of the tunneling rates and the tunneling conductance approximation. It well illustrates that the tunneling conductance approximation is unable to describe correctly the evolution of the tunneling rates with the bias voltage. Of course, the concept of tunneling conductance can be improved with more or less complex models[12,13] to take into account the bias dependence of the tunneling rates. Nevertheless the most important improvment offered by a full tunneling rate calculation *via* transfer Hamiltonian method lies in the parameters needed for the simulation: only physical fundamental data of the system are necessary and without fitting parameter.

In fact, provided we have access to the various electron wave-functions of the system, we are able to determine the tunneling rates. Thus, if a good model of the semiconductor quantum dots is used, including electron electrostatic interaction and energy quantization, the tunneling rates of semiconductor Coulomb blockade devices can be calculated. This is all the more interesting that tunneling conductance and orthodox theory should be unable to give an accurate description of these devices[5].

## C. Semiconducting Coulomb blockade devices

In previous articles[4,5] we have extensively describe a model based on Hartree method to calculate the wave-function of the electrons in a silicon (and more generally semiconductor) quantum dot for Coulomb blockade application. This model also takes into account the bias voltage applied along the quantum dot.

To calculate the various tunneling rates of a MI*Si*IM structure like the one presented in figure 4, we use two kinds of wave-functions: first, the analytical wave-functions of equations (57) describe the electrons inside the metallic electrodes ; then, the wave-function given by the numerical resolution of the Hartree Hamiltonian of Ref. 5. Using Bardeen's formula, we have access to the matrix element $M$ and consequently to the tunneling rates either for an electron coming from an electrode and arriving in the quantum dot

$$\Gamma_{\text{elec} \to \text{dot}} = \sum_{\mathcal{E}_{\text{dot}}} \frac{2\pi}{\hbar} |M_{\text{elec} \rightleftharpoons \text{dot}}|^2 \rho_{\text{elec}}(\mathcal{E}_{\text{dot}}) l_{\text{dot}}(\mathcal{E}_{\text{dot}}) f_{\text{elec}}(\mathcal{E}_{\text{dot}}), \tag{65}$$

or for an electron coming from the dot and going to the electrode

$$\Gamma_{\text{dot} \to \text{elec}} = \sum_{\mathcal{E}_{\text{dot}}} \frac{2\pi}{\hbar} |M_{\text{elec} \rightleftharpoons \text{dot}}|^2 \rho_{\text{elec}}(\mathcal{E}_{\text{dot}}) g_{\text{dot}}(\mathcal{E}_{\text{dot}}) \left[ 1 - f_{\text{elec}}(\mathcal{E}_{\text{dot}}) \right]. \tag{66}$$

In the two above expressions, $l_{\text{dot}}(\mathcal{E}_{\text{dot}})$ and $g_{\text{dot}}(\mathcal{E}_{\text{dot}})$ correspond, respectively, to the number of free states and to the number of electrons on the energy level $\mathcal{E}_{\text{dot}}$. Of course, it is practically impossible to sum over all possible states of the quantum dot: only a few number of excited states can be studied.

An example of tunneling rates evolution with the bias voltage for a 30 Å radius spherical quantum dot is presented in figure 7. As we can see, the variations of tunneling rates are far from being similar to those found for metallic devices. In particular the tunneling rates can decrease with the bias voltage. In fact this effect is due to the influence of bias potential on the electron density which concentrates near output barrier. This phenomenon, already discussed in Ref. 5 and which leads to the notion of negative differential conductance, will be more extensively reviewed in a future article.

From information on tunneling rates, it is now possible to deduce the $I(V)$ characteristics of Coulomb blockade devices. In this view two numerical algorithms could be employed: the master equation and the Monte-Carlo method.

## IV. FROM TUNNELING RATES TO CURRENT/VOLTAGE CHARACTERISTICS

### A. Master equation method Vs Monte-Carlo point of view: metallic case

A classical way to determine $I(V)$ characteristics lies on the master equation technique. This method is based on a balance of the probability density $P(N, t)$ of finding $N$ electrons in the island at time $t$. For instance, considering only the sequential transport of the electron one by one, the differential equation governing the MIMIM structure of figure 4 is given by

$$P(N, t+dt) = P(N,t)\left[[1-\Gamma_{L\to dot}(N)dt][1-\Gamma_{R\to dot}(N)dt][1-\Gamma_{Dot\to L}(N)dt][1-\Gamma_{dot\to R}(N)dt]\right]$$
$$+ P(N+1, t)\left[\Gamma_{dot\to R}(N+1)dt + \Gamma_{Dot\to L}(N+1)dt\right]$$
$$+ P(N-1, t)\left[\Gamma_{R\to dot}(N-1)dt + \Gamma_{L\to dot}(N-1)dt\right], \quad (67)$$

that is at the fist order

$$\frac{\partial P(N,t)}{\partial t} = P(N+1,t)\left[\Gamma_{dot\to R}(N+1) + \Gamma_{Dot\to L}(N+1)\right]$$
$$+ P(N-1,t)\left[\Gamma_{R\to dot}(N-1) + \Gamma_{L\to dot}(N-1)\right]$$
$$- P(N,t)\left[\Gamma_{L\to dot}(N) + \Gamma_{R\to dot}(N) + \Gamma_{Dot\to L}(N) + \Gamma_{dot\to R}(N)\right]. \quad (68)$$

By solving this differential equation system on a finite number of possible states (i.e. $N$ lies in a finite interval), we can determine the probability $P(N,t)$ as a function of time. The steady-state current in the structure is then calculated by counting the number of charge that cross one of the tunnel junction (either the left or right barrier)

$$I = -q\sum_N P(N)\left[\Gamma_{dot\to R}(N) - \Gamma_{R\to dot}(N)\right], \quad (69)$$

with $P(N) = P(N, t \to \infty)$.

The $I(V)$ characteristic obtained by the master equation method of the MIMIM structure of figure 5 is shown in figure 8a (continuous line). The value of the current is consistent with the first attempt of experimental current determination for single-nanocristal structures[20]. This is all the more remarkable because no high-level fit parameter (such as tunneling rates or tunneling conductance) is used. One of the consequence of using a full calculation of the tunneling rates is the non symmetrical behavior of the current as a function of voltage applied. Of course, the evolution of the current on the Coulomb plateau is different from the one obtained by the tunneling conductance method.

Nevertheless the master equation suffers from the fact that many states of the island must be considered in order to deduce the current. This is not capital in simple devices such as MIMIM structures but can become a major issue when studying complex devices with many islands (the number of total device states being in power of the number of island). One way to circumvent this limitation consists in using a Monte-Carlo method.

Indeed, the Monte-Carlo algorithm, consists in determining statistically, the evolution of the number of electrons in the island as a function of time. To do so, we first pick a random number $U$ uniformly distributed on the range [0, 1] to determine the free flight time after which a tunnel event occurs:

$$\tau = -\frac{\ln(1-U)}{\sum_i \Gamma_i}, \quad (70)$$

where $\sum_i \Gamma_i$ represents the sum of tunneling rates of all possible events. Then, at the time $t = t + \tau$, we choose randomly which tunneling event occurs by taking into account the respective probability of each tunnel event.

The current is directly calculated by taking the average value of the number of charges crossing a tunnel barrier per unit of time. The $I(V)$ characteristic obtained by this Monte-Carlo method is presented in figure 8a (cross symbol). As we can see there is a great dispersion in the values of current around the average value which is represented by the results of the master equation method. Perfect to solve non-stationary problem or the shot-noise induced by the sequential tunneling phenomenon, the Monte-Carlo algorithm suffers from its advantages. Thus with this powerful method, it seems difficult to give a global information such as the correct shape of $I(V)$ characteristic without particle noise.

To combine the advantages of the Monte-Carlo method with the ones of the master equation, we study the particle evolution as a function of time thanks to the "classical" Monte-Carlo algorithm described previously but we calculate the probability $P(N)$ of finding $N$ electrons in the island (by counting the meaning time $N$ electrons are present) in steady-state regime. The current is then calculated *via* equation (69). The $I(V)$ characteristic shown from this technique is exhibited in figure 8b and compared to the curve obtained by master equation: the results are similar and the Monte-Carlo method makes unnecessary the knowledge of all possible states of the system. As a consequence, we have at our disposal two kinds of Monte-Carlo techniques to determine $I(V)$ characteristics: on the one hand, a Monte-Carlo method coupled to a probabilistic calculation of the current to know the current/voltage general shape and on the other hand, a "classical" Monte-Carlo algorithm able to study in detail the non-stationary or shot-noise problems.

### B. $I(V)$ characteristic of semiconducting Coulomb blockade devices

Let's see as last application the use of these methods in the case of semiconducting Coulomb blockade device as the MI*Si*IM structure. Knowing the various tunneling rates presented in figure 7 and using the same Monte-Carlo method as the one explained in the previous subsection, we have access to current/voltage characteristics of the device.

For instance figure 9 shows the $I(V)$ characteristic of an Al—SiO$_2$—Si quantum dot—SiO$_2$—Al device with a 30 Å radius quantum dot at 30 K; the left and right barriers are respectively 15 Å and 12 Å thick (thickness taken from the edge of the sphere to the metallic plate); the electrode surface is 140 nm × 140 nm. These geometrical parameters coupled to the Al work function, the Al Fermi level energy, the electron effective mass and the Si/SiO$_2$ barrier height are the only input data of the simulation. First, the eigenstates (ground state and some excited levels) of the quantum dot embedded in SiO$_2$ are calculated as detailed in Ref. 5. Then, the transfer tunneling Hamiltonian method is used to derive the tunneling rates. Finally, $I(V)$ characteristic is deduced from the Monte-Carlo method. In figure 9 we observe a classical Coulomb blockade characteristic with two threshold voltages (at approximately 0.45 V and 0.79 V): we can notice that only one or two electrons can be stored simultaneously in this very small quantum dot. It should be underlined that the current flowing in the structure is particulary low. Except for the multidot memory device, such current is incompatible with a direct use of semiconducting Coulomb blockade devices in classical circuit architectures yet.

## V. CONCLUSION

With a view to developing a single electron device simulator, we have presented in this paper a set of method which can be applied in order to obtain $I(V)$ characteristics of Coulomb blockade devices from the knowledge of the physical fundamental parameters of the system. Either metallic or semiconducting structures can be treated. It is based on the weak coupling approximation of the theory of tunneling transfer Hamiltonian coupled with the Bardeen's formula. Thus tunneling rates could be calculated from the knowledge of the electron wave-function in the different part of the system. This wave-function is either given by an analytical approximation for the case of metallic electrodes or by a numerical resolution of the Poisson-Schrödinger equation in the Hartree approximation for the semiconductor quantum dot. Finally, $I(V)$ characteristics of the devices are deduced from a Monte-Carlo treatment.

Of course, simulation of Coulomb blockade devices from fundamental physical parameters is of first importance to predict and analyze experimental behavior. It could be used to guide the design of future single electron architecture. In particular, the evolution of tunneling rates which decreases with the bias voltage in semiconductor quantum dot, due to the deformation of the wave-function induced by the electric field, seems to announce that simple MI*Si*IM structure could present negative differential conductance effect. This point and a detailed study of $I(V)$ characteristic of semiconductor Coulomb blockade devices will be presented in a future article.

could be considered as an electrode) even if in this section we speak only of quantized energy level, generalization to continuum will be seen later.

[23] This is in fact the approximation of de J. R. Oppenheimer in his famous article of Ref. 21.

[24] $$\iiint_\mathcal{V} u\, \vec{\nabla} \left([M]^{-1}\vec{\nabla} v\right)\, dV = \oiint_\mathcal{S} u \cdot [M]^{-1} \vec{\nabla} v\, dS - \iiint_\mathcal{V} \vec{\nabla} u \cdot [M]^{-1} \vec{\nabla} v\, dS$$

[25] Nevertheless we should to neglect this term since $\langle \Phi_R^k | \Phi_L^0 \rangle$ is a second-order term.

[26] The lengthes $L_y$ and $L_z$ correspond to the size of the electrode in the direction perpendicular to the tunneling process

[27] This assumption is true if we neglect the variation of the electron effective mass along the $x$ axis in the Hamiltonians $\mathcal{H}_y$ and $\mathcal{H}_z$.

**List of Figures**

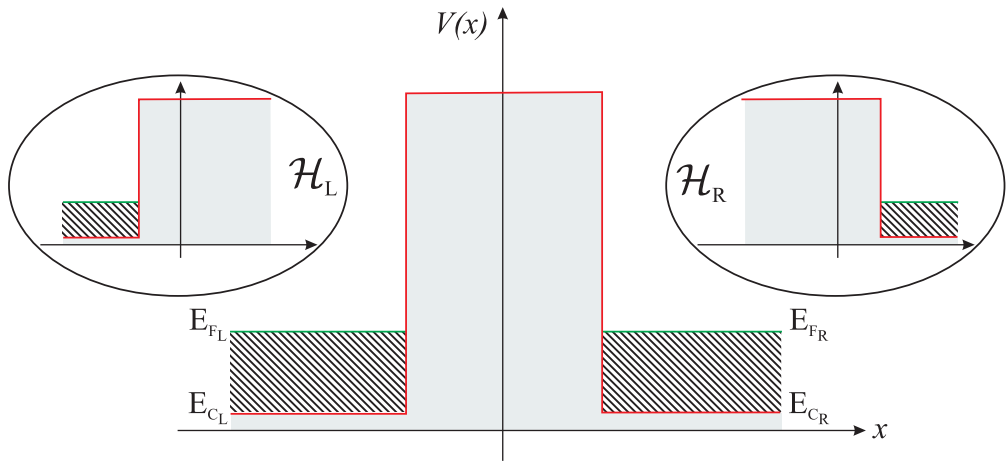

FIG. 1: Description of a single tunnel barrier between two isolated electrodes *via* the tunneling Hamiltonian approach: the system is decomposed into two independent sub-systems.

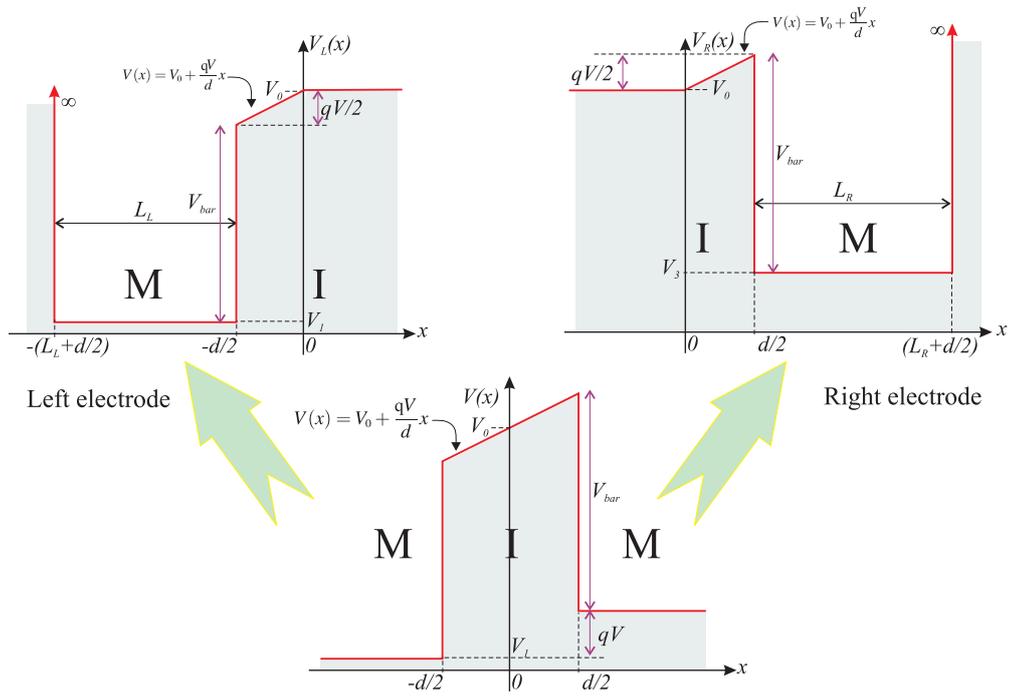

FIG. 2: Description of a single biased tunnel barrier potential and the two left and right Hamiltonian resulting from the decomposition described in section II.

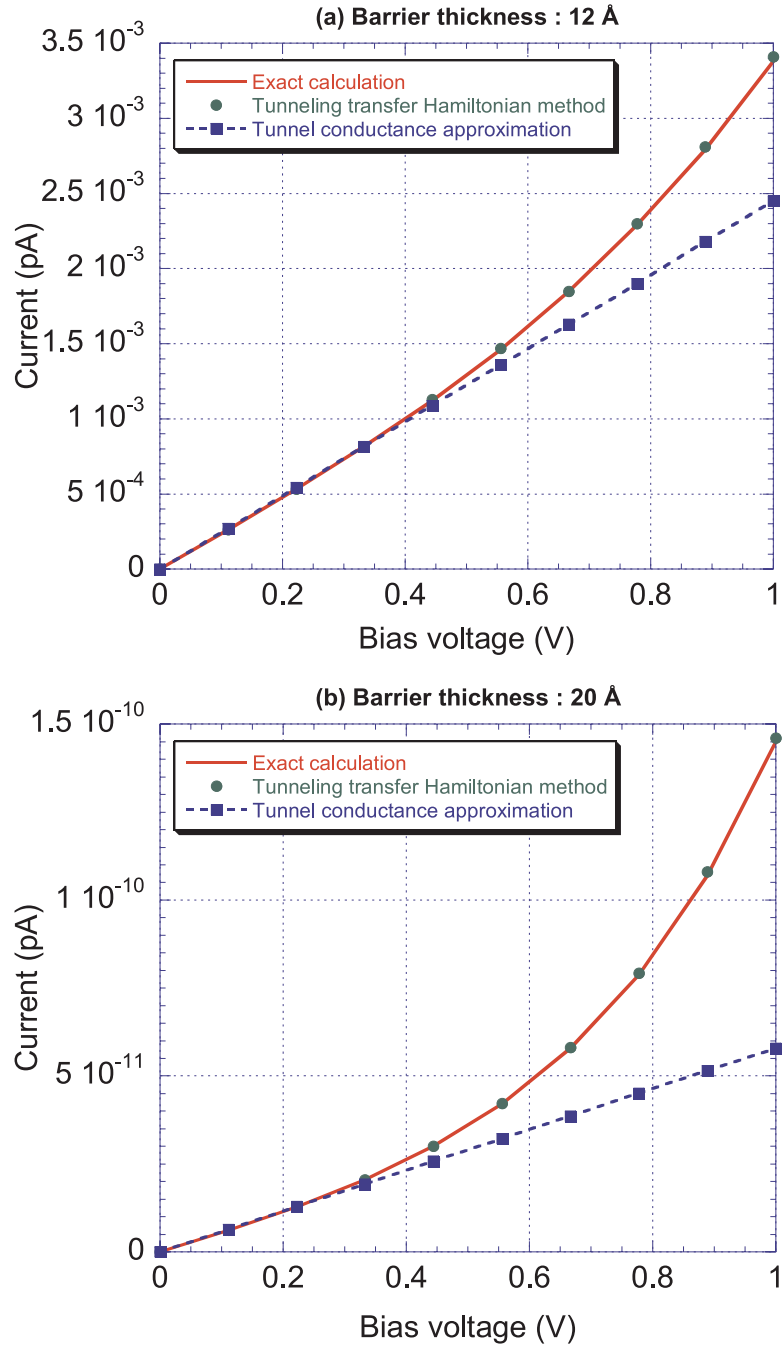

FIG. 3: Current/voltage characteristic $I(V)$ of the tunnel junction Gold–$SiO_2$–Gold for various barrier thicknesses: (a) 12 Å, (b) 20 Å. The value of the current results from three methods: exact calculation (continuous line), tunneling transfer Hamiltonian (symbol) and tunneling conductance approximation (dash line).

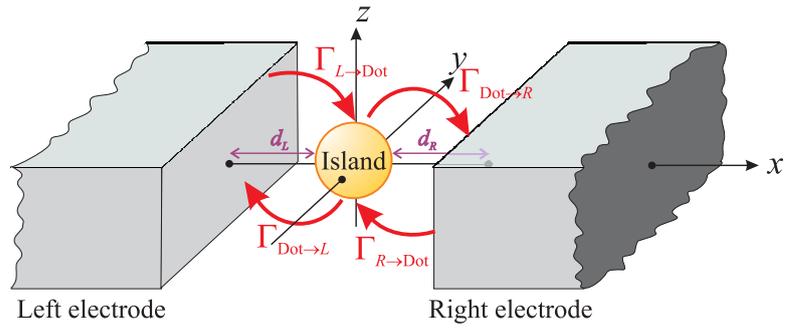

FIG. 4: Schematic of a typical Metal—Insulator—Island—Insulator—Metal structure ; the island can be either a semiconductor quantum dot or a Metallic cluster. The simulation of such devices requires the knowledge of four parameters: the four tunneling rates $\Gamma_{\text{dot}\to R}$, $\Gamma_{\text{dot}\to L}$, $\Gamma_{R\to\text{dot}}$, $\Gamma_{L\to\text{dot}}$.

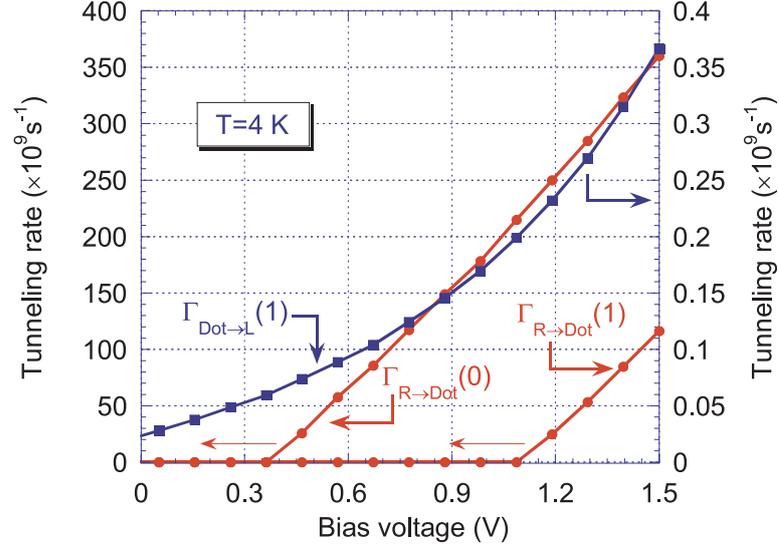

FIG. 5: Evolution of tunneling rates as a function of bias voltage for a Al—SiO$_2$—Al—SiO$_2$—Al structure (left barrier thickness: 23 Å, right barrier thickness: 23 Å island surface along $x$ axis $L_y L_z = 15$ nm$^2$ ). $\Gamma_{R \to dot}(N)$ represents the tunneling rate from the right electrode to the island where $N$ electrons are stored and $\Gamma_{dot \to R}(1)$ is the tunneling rate from the island containing one electron to the left electrode.

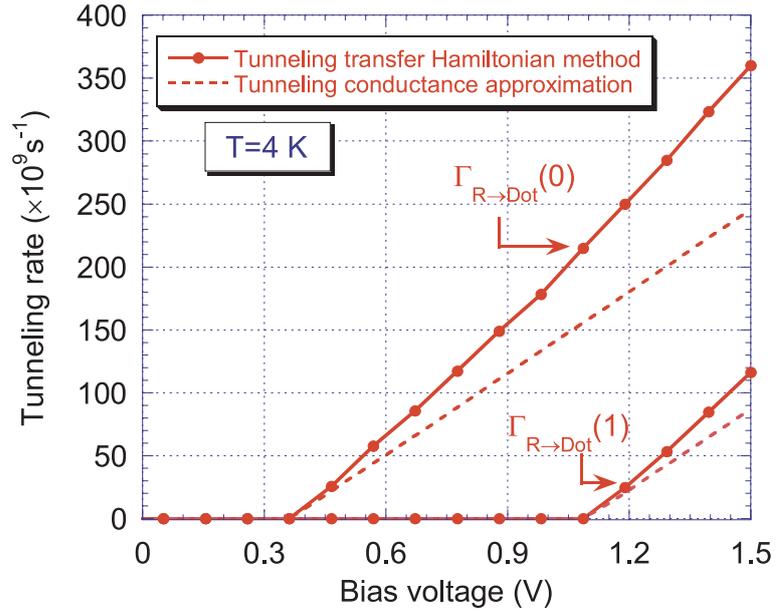

FIG. 6: For the same simulation parameter as in figure 5, comparison between a full calculation of the tunneling rates thanks to the tunneling transfer Hamiltonian (continuous line) and the tunneling conductance approximation (dash line).

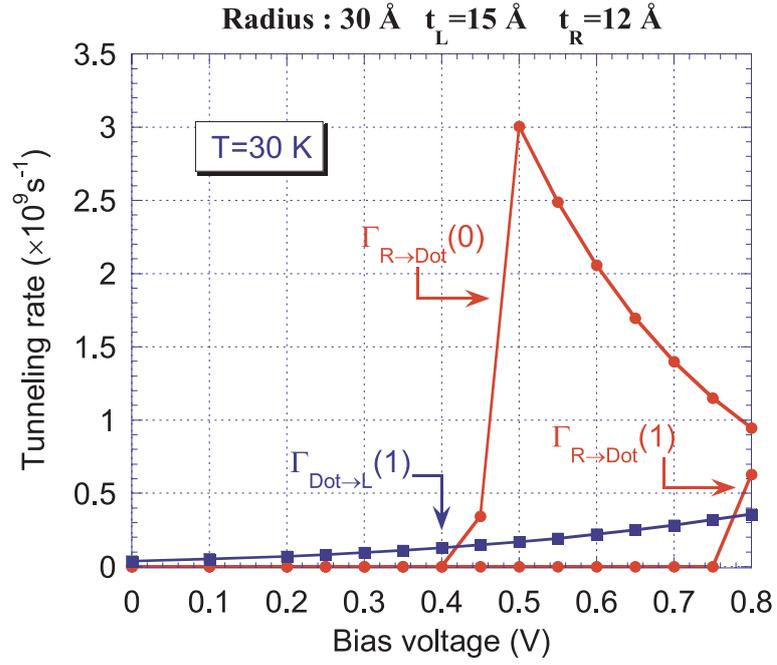

FIG. 7: Evolution of tunneling rates as a function of bias voltage for a Al—SiO$_2$—Si quantum dot—SiO$_2$—Al structure (left barrier thickness: 15 Å, right barrier thickness: 12 Å, quantum dot radius: 30 Å ). $\Gamma_{R \to dot}(N)$ represents the tunneling rate from the right electrode to the island where $N$ electrons are stored and $\Gamma_{dot \to R}(1)$ is the tunneling rate from the island containing one electron to the left electrode.

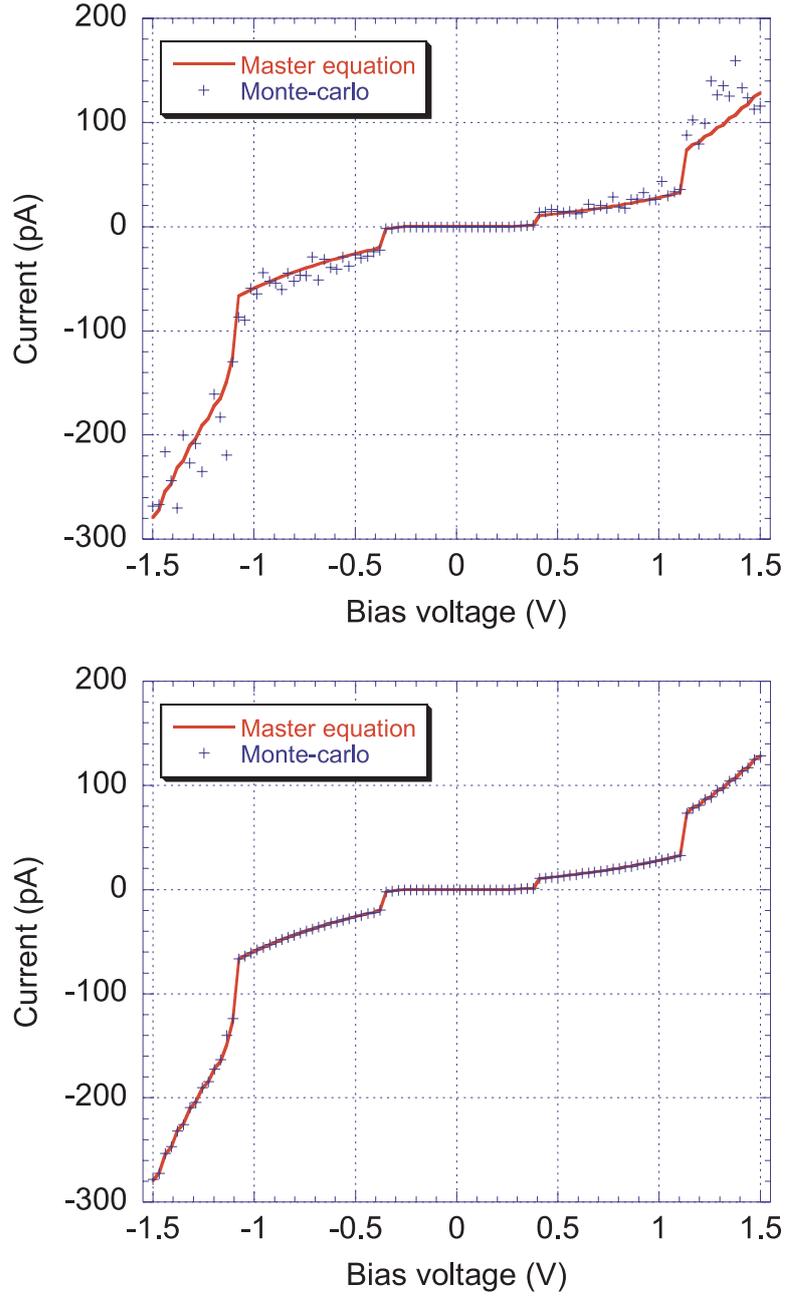

FIG. 8: Comparison of $I(V)$ characteristics of the MIMIM structure presented in figure 5 (a) between the master equation and the Monte-Carlo method and (b) between the master equation and the improved Monte-Carlo method.

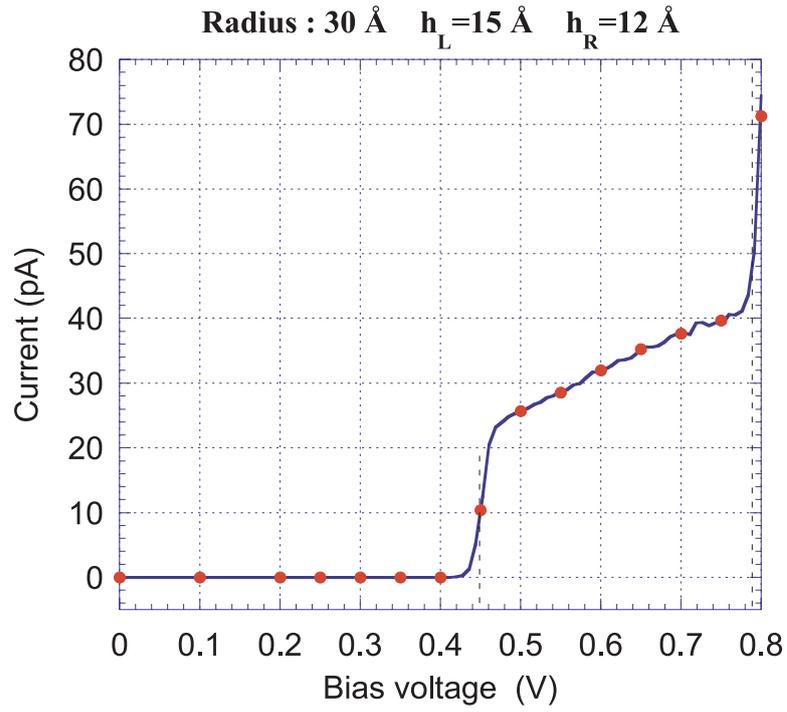

FIG. 9: $I(V)$ characteristics for a MI$Si$IM structure with a 30 Å radius silicon spherical quantum dot at 30 K. The left and right tunnel junctions are 15 Å and 12 Å thick respectively and the electrodes are made of aluminium.